%% file: main.tex
\newif\ifraw   % Schalter f"ur \tt-Stil (true schaltet auf \tt-Stil
\rawfalse
\documentclass[11pt,fleqn,leqno]{article}
\usepackage{amsmath}
\usepackage{amssymb}
\usepackage{amsthm}
\catcode`\@=11
\catcode`\@=12
\usepackage{enumerate}
\newcommand{\zeile}{\hfill\\}
\newcommand{\komma}{\quad ,}
\newcommand{\punkt}{\quad .}
\swapnumbers
\theoremstyle{plain}
\newtheorem{theorem}{Theorem}[section]
\newtheorem*{theorem*}{Theorem}
\newtheorem{proposition}[theorem]{Proposition}
\newtheorem*{proposition*}{Proposition}
\newtheorem{lemma}[theorem]{Lemma}
\newtheorem*{lemma*}{Lemma}
\newtheorem{corollary}[theorem]{Corollary}
\newtheorem*{corollary*}{Corollary}
\theoremstyle{definition}
\newtheorem{definition}[theorem]{Definition}
\newtheorem*{definition*}{Definition}
\theoremstyle{remark}
\newtheorem{remark}[theorem]{Remark}
\newtheorem*{remark*}{Remark}
\newtheorem{note}[theorem]{Note}
\newtheorem*{note*}{Note}

\newtheorem*{cit*}{Citation}
\newcommand{\ein}{\hangindent1cm\hangafter1\zeile}
\newcommand{\aus}{\hangindent0cm\hangafter1\zeile}
\newcommand{\RR}{\ensuremath{\mathbb{R}}}   % Symbol für reelle Zahlen
\newcommand{\NN}{\ensuremath{\mathbb{N}}}   %       nat"urliche Zahlen
\renewcommand{\d}{\partial} % partieller Differentialoperator
\newcommand{\norm}[1]{\lVert#1\rVert}
\newcommand{\abs}[1]{\lvert#1\rvert}
\newcommand{\absbig}[1]{\left\lvert#1\right\rvert}
\newcommand{\hnorm}[2]{\lVert#1\rVert_{H^{#2}}}
\newcommand{\cnorm}[2]{\lVert#1\rVert_{C^{#2}}}
\newcommand{\lnorm}[2]{\lVert#1\rVert_{L^{#2}}}

\DeclareMathOperator{\tr}{tr}
\DeclareMathOperator{\id}{id}

\DeclareMathOperator{\im}{Im}
\DeclareMathOperator{\diam}{diam}

\DeclareMathOperator{\ind}{ind}

\newcommand{\leftsuper}[1]{{}^{\scriptscriptstyle #1}}
\newcommand{\viernabla}{\leftsuper{4}\nabla}
\newcommand{\vierR}{\leftsuper{4}\!R}
\newcommand{\wt}{w(\tau)}
\newcommand{\wj}{w^j}
\newcommand{\wjm}{w^{j-1}}
\newcommand{\wjp}{w^{j+1}}
\newcommand{\dtau}{\d_{\tau}}
\newcommand{\Dalpha}{D^{\alpha}}
\newcommand{\Aij}{A^i(w^{j},N^{j})}
\newcommand{\Aijm}{A^i(w^{j-1},N^{j-1})}
\newcommand{\Bj}{B(w^{j},N^{j},DN^{j})}
\newcommand{\Bjm}{B(w^{j-1},N^{j-1},DN^{j-1})}
\newcommand{\Nj}{N^j}
\newcommand{\Nt}{N(\tau)}
\newcommand{\Njm}{N^{j-1}}
\newcommand{\Njp}{N^{j+1}}
\newcommand{\Njt}{N^j(\tau)}

\newcommand{\Ltjm}{L_{\tau}(\wjm,D\wjm)}
\newcommand{\Ltj}{L_{\tau}(\wj,D\wj)}
\newcommand{\Lj}{L(\wj,D\wj)}
\renewcommand{\dj}[1]{\delta_j#1}
\newcommand{\djm}[1]{\delta_{j-1}#1}
\newcommand{\djmm}[1]{\delta_{j-2}#1}
\newcommand{\dt}[1]{\delta_{\tau}#1}
\ifraw
\input wiggly.tex  % fuer wavy underlining
\fi
\title{Local Prescribed Mean Curvature foliations in cosmological spacetimes}
\author{Oliver Henkel\thanks{Present address: Heinrich--Hertz--Institut
    f\"ur Nachrichtentechnik Berlin GmbH, Einsteinufer 37, 10587 Berlin,
    Germany}\\ 
  Max Planck Institute for Gravitational Physics\\ Am M\"uhlenberg 1\\
  14476 Golm, Germany}  
\date{August 01, 2001}
\begin{document}
\maketitle
\begin{abstract}
    A theorem about local in time existence of spacelike foliations with
    prescribed mean curvature in cosmological spacetimes will be
    proved. The time function of the foliation is geometrically
    defined and fixes the
    diffeomorphism invariance inherent in general foliations of spacetimes.
    Moreover, in contrast to the situation of the more special
    constant mean curvature foliations, which play an important role in
    global analysis of spacetimes, this theorem overcomes the existence
    problem arising from topological restrictions for surfaces of constant
    mean curvature.
\end{abstract}
\tableofcontents
%
% Umschaltung auf \tt-Stil?
\ifraw
\renewcommand{\emph}{\underline}
\renewcommand{\textbf}{\underwiggle} 
\renewcommand{\baselinestretch}{2}
\tt
\fi
\include{intro}
% 
\include{basics}
%
\include{body}

%
\include{conclusion}
%
\include{bib}
%
%
%
%
%
\end{document}

%% file: intro.tex
\section{Introduction}
Spacelike foliations arise in many contexts in General
Relativity. This is related to the fact, that the world around
us seems three dimensional and we are used to think about dynamics as a
sequence of processes parametrized by some time function. But in Relativity
there is no canonical time coordinate, thus any time function only fixes 
(a part of) the diffeomorphism invariance of the 
theory but this gauge fixing is a priori arbitrary unless the time function
is tied to some geometrically defined quantity. For that reason, an
arbitrary time function does not provide any information about the global
structure of the spacetime under consideration. For example, any thin,
spacelike strip in Minkowski spacetime can be regarded as global in time,
by stretching the time coordinate with an appropriate diffeomorphism. This
problem is connected with the lack of a canonical metric background
structure for Einstein's field equations. 

In this work I will define and prove a uniqueness and local in time
existence theorem for a spatially global, geometrically defined time
coordinate in cosmological spacetimes in terms of a 
\emph{Prescribed Mean Curvature (PMC)} foliation, defined as follows: The
leaves $\{S_t\}$ are determined implicitly by the requirement, that given
an initial Cauchy surface $S_0:=\Sigma$ 
subject to a certain condition, the mean curvature at each point 
$p_t \in S_t$ is defined by the relation 
$H(p_t)=H(p_0)+t$, where $p_t$ and $p_0 \in S_0$ are connected by the
integral curves $\gamma(t)$ of the normal vector field of the leaves. 
Therefore, the elements in this construction are purely
geometrical in nature, the time function obtained by this procedure
depends only on the choice of the initial Cauchy surface and we get a
geometrical measure of the (timelike) size of its Cauchy development as
long as this time coordinate exists. I will refer to the leaves of such a
foliation as surfaces of prescribed mean curvature. 
Note, that there is another definition for surfaces with prescribed mean
curvature in the literature ('H-surfaces'), compare \cite{g}. But unlike the 
definition given there, where the spacetime and a function H prescribing the
mean curvature are given quantities, the definition here relates to the PMC
foliation as a whole, providing an intrinsic prescription 
without having given the spacetime or a function $H$ in advance: 
Coupling the PMC equations to Einstein's evolution equations with matter  
(such that the Cauchy problem in harmonic coordinates is well posed),
one gets a system of equations, which constructs a
spacetime foliated in this way, where the mean curvature 'develops' from
leaf to leaf. 

Given this kind of foliation globally, we have a tool at hand for the
global analysis of spacetimes in an invariant manner, in terms of an
asymptotic analysis of a system of 
partial differential equations, throwing away the differential geometric
difficulties arising from the diffeomorphism invariance. For an
introduction to global issues in Relativity compare the survey article 
\cite{an}. Related topics are the development of singularities, the cosmic
censorhip conjecture and the closed universe recollapse
conjecture. Introductions to these topics can be found in \cite{me} (global
existence and cosmic censorship) and \cite{bt},\cite{bgt} (closed universes
and recollapse). 

A related construction are the
well-known constant mean curvature (CMC) foliations, see the survey article
\cite{r} or \cite{mt} for reference. Obviously they are a
special case of a PMC foliation, when the initial Cauchy surface itself has
constant mean curvature, and this fact sheds some light about the origins
of PMC foliations. Indeed, the need for such a generalization arises from the
fact, that there are strong topological restrictions concerning the
existence of even a single CMC hypersurface. There exist spacetimes, which
do not possess any constant mean curvature hypersurface at all, see \cite{ba}. 
Conequently, many of the results obtained by constructions
with CMC foliations presuppose the existence of at least a single CMC
hypersurface, which then has to be verified in each concrete case. Contrary
to this, for the existence of a local in time PMC foliation there is a
simple and easily verified criterion (compare theorem \ref{thm.pmc}),
leading to a broad class of spacetimes, possessing at least a local in time
PMC foliation. The similarity between PMC and CMC foliations then gives rise
to the hope, that the global results obtained in the CMC cases (under the
additional existence assumptions mentioned above) can also be proved using
PMC foliations, an issue a forthcoming paper will be concerned with.

The organisation of this work is as follows. 
The section \ref{s.preliminaries} fixes notation and introduces
the basic equations and results used throughout this work. Its first part
is concerned with General Relativity, while the second part is devoted to
partial differential equations. Section \ref{s.localpmc} then contains the
construction of the local in time PMC foliation. In the last section the
main result will be stated and discussed.
% 
%
%
%
%

%%% Local Variables: 
%%% mode: latex
%%% TeX-master: "main"
%%% End: 

%% file: basics.tex
%
%
%
%
%
\section{Preliminaries}\label{s.preliminaries}
\subsection{Spacetimes and foliations}\label{s.spacetimes}
A spacetime is a pair $(M,g)$, where $M$ denotes a four dimensional
smooth and orientable Lorentz manifold with metric $g$ and signature
$(-+++)$. The metric induces structures like the Levi-Civita
connection $\viernabla$ and the curvature on $M$. 
The sign convention for the curvature is fixed by the definition 
$\vierR(X,Y)Z := \viernabla_X\viernabla_Y Z - \viernabla_Y\viernabla_X Z 
               - \viernabla_{[X,Y]} Z$, 
where $X$, $Y$ $Z$ are vector fields. The curvature tensor is then defined
as $\vierR(W,X,Y,Z) := g(W, \vierR(X,Y)Z)$ with Ricci tensor 
$\vierR_{\alpha\beta}=\vierR^{\mu}_{\alpha\mu\beta}$
and scalar curvature $\vierR = \vierR^{\mu}_{\mu}$, written in abstract
index notation of the Ricci calculus. The Einstein tensor reads 
$G_{\alpha\beta} = \vierR_{\alpha\beta} 
                 - \tfrac{1}{2}\vierR g_{\alpha\beta}$ 
and we can write Einstein's field equation as
\begin{subequations}
\begin{gather}
   G_{\alpha\beta} = 8\pi T_{\alpha\beta} \\
   \intertext{or equivalently (by $\vierR = -8\pi \tr T$)}
   \vierR_{\alpha\beta} = 8\pi 
     \left( T_{\alpha\beta} - \tfrac{1}{2}(\tr T)\,g_{\alpha\beta} \right)
   \komma
\end{gather}
\end{subequations}
where $T_{\alpha\beta}$ denotes the energy momentum tensor of the matter
fields. It is a symmetric tensor on $M$ with vanishing divergence as
a consequence of the Bianchi identities, a requirement which imposes
supplementary conditions on the matter fields coupled to the field
equation.\zeile 
The components of the energy momentum tensor with respect to an
observer, represented by a unit timelike vector $n$, have physical
meaning: Denote by 
$h_{\alpha\beta} := g_{\alpha\beta} + n_{\alpha}n_{\beta}$ the
orthogonal projector on $\{n\}^{\perp}$ in covariant notation, then we can
define the energy density, momentum density and the stress tensor by
\begin{subequations}
   \begin{gather}
      \rho := T_{\mu\nu}n^{\mu}n^{\nu} \\
      j_{\beta} := -T_{\mu\nu}n^{\mu}h^{\nu}_{\beta} \\
      S_{\alpha\beta} := T_{\mu\nu}h^{\mu}_{\alpha}h^{\nu}_{\beta}
   \punkt
   \end{gather}
\end{subequations}

In this work we confine ourselves to \emph{cosmological} solutions $(M,g)$
of Einstein's field equations. Due to \cite{ba} these are globally
hyperbolic and spatially compact spacetimes, where the Ricci 
tensor contracted twice with any timelike vector is non-negative (timelike
convergence condition). This last condition can be reexpressed in terms of
the matter variables as $\rho+\tr S \ge 0$ for any observer, known as the
strong energy condition.\zeile
An introduction to General Relativity with a treatment of  
the Cauchy problem for the field equations is \cite{w1}, while a 
recent presentation with a deeper analysis can be found in 
\cite{fr}.

Now let us pay attention to an additional structure. 
A foliation $\{S_t\}$, $t \in I \subset \RR$ ($I$ interval containing zero)
of (a part of) $(M,g)$ by spacelike hypersurfaces induces on 
each leaf the unit normal vector field $n$, the metric 
$h_{\alpha\beta} = g_{\alpha\beta} + n_{\alpha}n_{\beta}$, which also 
serves as orthogonal projection and the second fundamental form
$k_{\alpha\beta}:=-h_{\alpha}^{\mu}h_{\beta}^{\nu}\,\viernabla_{\mu}n_{\nu}$
(the definition of $k_{\alpha\beta}$ fixes the sign conventions used in this
work). The second fundamental form is a symmetric tensor, intrinsic to the
leaves of the foliation, and it can also be written as the
Lie derivative of the 
3-metric $h$ with respect to the normal vector field, 
$k_{\alpha\beta}=-\tfrac{1}{2}{\cal L}_n h_{\alpha\beta}$.
The 3-metric determines further geometrical objects on the leaves,
such as the Levi-Civita connection $\nabla$ and the curvature tensor
$R(\cdot)$. Tensors intrinsic to the leaves of the foliation will carry
Latin indices in the abstract index notation.\zeile 
The parameter $t$ of the foliation has timelike gradient and thus can be
regarded as (coordinate-) time. Given
local coordinates $(x^i)$ on $S_0$, we can Lie-transport them to
neighbouring leaves along an arbitrary family of
transversal curves, parametrized by $x$. We will express equations
containing coordinate components with respect to the adapted
coordinates $(t,x)$.\zeile  
We define the lapse function $N$ and the shift vector $\nu \perp n$ on
the leaves by the formula
\begin{equation}
   \d_t = Nn + \nu
   \quad\text{, thus}\quad
   \begin{gathered}
      N = -g(\d_t,n) \\
      \nu = \d_t - Nn
   \end{gathered}
\end{equation}
Then we find $1 = dt(\d_t) = N\,dt(n)$. Further, $dt$ is
(co-)orthogonal on the leaves and if we denote the conormal of
the leaves by $\sigma$ we see that 
$dt = -N^{-1}\sigma$ or $\sigma = -N dt$.
Thus $N^{-1}$ measures the length of $dt$, which 
can be interpreted as follows: For an observer along $n$ $N$ measures
the elapsed proper time from $S_t$ to $S_{t+dt}$ or equivalently, the
elapsed coordinate time $dt$ along a journey along $n$ between two
leaves, separated by a unit distance (measured along $n$) is $N^{-1}$, 
thus $N^{-1}$ measures the elapsed coordinate time along $n$. 
Note that these relations have only infinitesimal meaning and for the
elapsed proper time along $n$ between $S_{t_1}$ and $S_{t_2}$ we have
$\int_{t_1}^{t_2}N$.

The coordinates adapted to the foliation together with lapse and shift fix
the diffeomorphism invariance and we get a 3+1-split of the field equations
consisting of constraint equations intrinsic to the leaves and evolution
equations.\zeile
The constraint equations are a consequence of the equations of Gauss and
Codazzi, where Einstein's equations have been used to eliminate the
curvature terms of the 4-geometry. The result is
\begin{subequations}
\begin{gather}
   R + H^2 - \abs{k}^2 = 16 \pi \rho 
   \qquad\,\text{(Hamiltonian constraint)}\\
   \nabla^j k_{ij} - \nabla_i H = 8\pi j_i
   \qquad\quad\!\text{(momentum constraint)}
   \komma
\end{gather}
\end{subequations}
with $\abs{k}^2 = k_{\alpha\beta}k^{\alpha\beta}$ and $H=\tr k$
denotes the mean curvature of the leaves.\zeile
The remaining evolution equations for the components of the first and
second fundamental forms are known as ADM equations. They read in adapted
coordinates 
\begin{subequations}\label{e.adm}
\begin{gather}
   \d_t h_{ij} = -2Nk_{ij} + \nabla_i\nu_j + \nabla_j\nu_i \\
   \d_t k_{ij} = -\nabla_i\nabla_j N
                   + N \Big(
                         R_{ij} + Hk_{ij} - 2k_i^r k_{rj} -8\pi
                         (\, S_{ij}+\tfrac{1}{2}(\rho-\tr S)h_{ij} \,)
                      \Big) \\
   \qquad\quad\,\, + \nu^r\nabla_r k_{ij} 
                   + k_{rj}\nabla_i\nu^r + k_{ir}\nabla_j\nu^r 
                   \notag
   \komma
\end{gather}
\end{subequations}
where the first equation is merely a rewriting of the definition of the
second fundamental form, $k=-\tfrac{1}{2}{\cal L}_n h$. Taking the
trace of the second equation and eliminating the scalar curvature $R$ by
the Hamiltonian constraint we obtain the lapse equation
\begin{equation}\label{e.lapse}
   \Delta N + N \Big(
                   \abs{k}^2 + 4\pi (\, \rho + \tr S \,)
                \Big)
     = (\d_t -\nu) H
   \komma
\end{equation}
which serves as a constraint of the foliation, induced by Einstein's
equations. Note, that in cosmological spacetimes, the term in brackets is
always non-negative.
\subsection{Functional analysis and partial differential equations}
At first I state some basic functional analytic definitions. 
Making use of the multi index notation, the Sobolev spaces are defined as
\begin{gather*}
   W^{k,p}(\RR^n) := \{f \in L^p(\RR^n) \;|\;
   \d^{\alpha}f \in L^p(\RR^n)\, \forall_{\alpha} \abs{\alpha} \le k\}
   \komma \\
   H^k(\RR^n) := W^{k,2}(\RR^n) \\
   \intertext{and for $s \in \RR$ we define}
   H^s(\RR^n) := \{ f \in {\cal S}'(\RR^n) \;|\;
   (1+\abs{\xi}^2)^{s/2} {\cal F}f \in L^2(\RR^n) \}
   \komma
\end{gather*}
where ${\cal S}'$ denotes the space of tempered
distributions and ${\cal F}$ is the Fourier
transform. There are continuous embeddings 
$H^s(\RR^n) \subset H^{s'}(\RR^n)$ for $s' < s$, thus the elements of
$H^s(\RR^n)$, $s \ge 0$ are functions. Furthermore, to compare different
Sobolev spaces I cite \cite{ra}, (Appendix A) where 
some basic facts from the theory of interpolation spaces are stated:\zeile
For $s_0 \neq s_1$ in $\RR$ let 
$s_{\theta} := (1-\theta)s_0 + \theta s_1$, $0 < \theta < 1$. Then 
\begin{equation*}
   [H^{s_0}(\RR^n),H^{s_1}(\RR^n)]_{\theta} = H^{s_{\theta}}(\RR^n)
   \komma
\end{equation*}
thus $H^{s_{\theta}}(\RR^n)$ is an intermediate space between
$H^{s_0}(\RR^n)$ and $H^{s_1}(\RR^n)$.

For the convenience of the reader I now list some (well known) inequalities,
that are central to the 
analysis of partial differential equations. I adopt the convention of
denoting any generic constant by $C$.

The perhaps most important estimate, we will make use of very often is
Gronwall's inequality. 
\begin{proposition}[Gronwall's inequality]\label{prop.gronwall}\ein
   Let $I \subset \RR$ be an interval, $t_0 \in I$ and
   $\alpha, \beta, u \in C(I,\RR_+)$, with
   \begin{equation*}
      u(t) \le \alpha(t) 
           + \absbig{\int_{t_0}^t \beta(s)u(s)\,ds }
   \end{equation*}
   for all $t \in I$.\zeile
   Then 
   \begin{equation*}
      u(t) \le \alpha(t)
           + \absbig{ \int_{t_0}^t \alpha(s)\beta(s)\:
                      e^{\abs{\int_s^t \beta(r)\,dr}}\: ds}
   \end{equation*}
   holds for all $t \in I$.
\end{proposition}
The proof of Gronwall's inequality in this particular form can be found 
in \cite{a}. 

The second important estimate, relating differentiability classes to each
other is the collection of the two Sobolev inequalities, which can be found
in standard textbooks about partial differential equations or functional
analysis, see for example \cite{rr}. Here, $D^s$
abbreviates the vector with components $D^{\alpha}$ for all $\alpha$ with
$s=\abs{\alpha}$.
\begin{proposition}[Sobolev inequalities]\ein
   {\rule{0pt}{1ex}}\zeile
   {\rm (S1)} $1 \le kp < n$
      \begin{gather*}
         W^{k,p}(\RR^n) \hookrightarrow L^{\frac{np}{n-kp}}(\RR^n) \\
         \lnorm{u}{\frac{np}{n-kp}} \le C \norm{u}_{W^{k,p}}
      \end{gather*}
   {\rm (S2)} $kp > n$
      \begin{gather*}
         W^{k,p}(\RR^n) \hookrightarrow L^{\infty}(\RR^n) \cap C^0(\RR^n)\\
         \lnorm{u}{\infty} \le C \norm{u}_{W^{k,p}}
      \end{gather*}
\end{proposition}
In particular we will use a special case of the second inequality,
\begin{corollary}[Sobolev embedding theorem]\label{cor.sobolev}\ein
   {\rule{0pt}{1ex}}\zeile
   $k > \tfrac{n}{2} + l$
      \begin{gather*}
         H^k(\RR^n) \hookrightarrow L^{\infty}(\RR^n) \cap C^l(\RR^n)\\
         \cnorm{u}{l} \le C \hnorm{u}{k}
      \end{gather*}
\end{corollary}

For the analysis of non-linear equations the Moser estimates are essential
\begin{proposition}[Moser inequalities]\label{prop.moser}\ein
   {\rule{0pt}{1ex}}\zeile
   {\rm (M1)} $f,g \in H^s(\RR^n) \cap L^{\infty}(\RR^n)$, 
      $s=\abs{\alpha}$
      \begin{equation*}
         \lnorm{D^{\alpha}(fg)}{2} \le C 
         \left(
           \lnorm{f}{\infty}\lnorm{D^s g}{2}
         + \lnorm{D^s f}{2}\lnorm{g}{\infty}
         \right)
      \end{equation*}
   {\rm (M2)} $f \in H^s(\RR^n) \cap W^{1,\infty}(\RR^n)$,
     $g \in H^{s-1}(\RR^n) \cap L^{\infty}(\RR^n)$, $s=\abs{\alpha}$
     \begin{equation*}
        \lnorm{D^{\alpha}(fg) - f D^{\alpha} g}{2} \le C
        \left(
          \lnorm{D^s f}{2} \lnorm{g}{\infty} 
        + \lnorm{Df}{\infty} \lnorm{D^{s-1}g}{2}
        \right)
     \end{equation*}
   {\rm (M3)} $f \in H^s(\RR^n) \cap L^{\infty}(\RR^n)$, 
     $F \in C^{\infty}(\RR)$, $F(0)=0$, 
     $B_R := \{x \in \RR^n \,|\, \norm{x} \le R \}$
     \begin{equation*}
        \lnorm{D^s F(f)}{2} \le c
          \left( \lnorm{f}{\infty} \right) \lnorm{D^s f}{2}
        \komma
     \end{equation*}
     \hspace{2em} with 
     $c(\lnorm{f}{\infty}) = C \norm{F}_{C^s(B_{\lnorm{f}{\infty}})}$ 
\end{proposition}
(see for example \cite{ra}).

Let us turn now to the analysis of special classes of partial
differential equations. We first consider hyperbolic equations:
\begin{definition}[Symmetric hyperbolic system]\ein
   Let $\{0\} \subset I \subset \RR$ denote an interval and  
   $G \subset \RR^k$ an open set. Further, for $i=1,\dots,n$ let $A^0,A^i,B$
   denote smooth functions on $I \times \RR^n \times G$, where $A^0-\id$,
   $A^i$ and $B$ have compact support. $A^0,A^i$ are assumed to take values
   in $\RR^{k \times k}$ and $B$ in $\RR^k$. Then we call the system
   \begin{equation*}
      A^0(t,x,u)\, \d_t u + A^i(t,x,u)\, \d_i u + B(t,x,u) = 0
   \end{equation*}
   for the unknown $u: I \times G \longrightarrow \RR^k$ 
   \emph{(quasilinear) symmetric hyperbolic}, if the matrices $A^0,A^i$ are
   symmetric and $A^0$ is uniformly positive definite.\zeile
   The system is called \emph{(inhomogeneous) linear symmetric hyperbolic},
   if it has the form
   \begin{equation*}
      A^0(t,x)\, \d_t u + A^i(t,x)\, \d_i u + B(t,x)u = f(t,x)
      \punkt
   \end{equation*}
\end{definition}
The following proposition holds (see e.g. \cite{ra} for a proof):
\begin{proposition}[Existence and Uniqueness of solutions]\ein
   \label{prop.standardenergy}
   \begin{enumerate}
   \item Let $u_0 \in C^{\infty}_0(\RR^n)$. Then there exists a unique
      solution $u \in C^{\infty}\left( [0,\infty[\, \times \RR^n \right)$
      of the linear symmetric hyperbolic system, with $u(0,x)=u_0(x)$.
      Moreover, $u$ satisfies the energy estimate
      \begin{equation*}
         \hnorm{u(t)}{s}^2 \le C \left(
           \hnorm{u(0)}{s}^2 + 
           \sup_{0 \le t' \le t}\hnorm{f(t')}{s}^2
         \right)
      \end{equation*}
   \item Let $s > \tfrac{n}{2}+1$, $u_0 \in H^s(\RR^n)$ with compact
      support. Then there is a 
      $T>0$ which depends on $\hnorm{u_0}{s}$ and a unique classical
      solution $u \in C^1([0,T] \times \RR^n)$ 
      of the quasilinear symmetric hyperbolic system with $u(0,x)=u_0(x)$
      and 
      \begin{equation*}
         u \in  \bigcap_{\NN \ni k < s} 
           C^k\left([0,T],H^{s-k}(\RR^n)\right)  
         \punkt
      \end{equation*}
      Moreover the energy estimate
      \begin{equation*}
         1 + \hnorm{u(t)}{s}^2 
           \le \left( 1+\hnorm{u(0)}{s}^2 \right)\:
               e\Big.^{C \textstyle \int_0^t 
                       \left(1+\cnorm{u(t')}{1}\right)\,dt'
                      }  
      \end{equation*}
      holds, providing some extension criterion if $\cnorm{u(t)}{1}$
      remains bounded in $[0,T]$.
   \end{enumerate}
\end{proposition}

In contrast to hyperbolic equations the study of elliptic operators differs
substantially from that of hyperbolic 
differential operators, because there is no finite propagation speed, not
even propagation at all. Therefore we cannot localize and have to analyse
elliptic equations as a whole instead.\zeile
In this work we need some standard facts about scalar linear elliptic
equations on 
compact manifolds. Basics about elliptic operators can be found for example
in \cite{rr} and some more advanced results on Fredholm theory of elliptic
operators on compact manifolds are presented in Cantor's article \cite{c}. 
In particular we find for the Laplacian 
$\Delta := -h^{ij}\nabla_i\nabla_j$ 
of a smooth compact Riemannian manifold $(\Sigma,h)$:
\begin{proposition}\ein
   Let $k \ge 0$ be some integer. 
   $\Delta : H^{k+2}(\Sigma) \longrightarrow H^{k}(\Sigma)$
   is a Fredholm operator with $\ind(\Delta)=0$,  
   $H^{k} = \ker(\Delta) \oplus \im(\Delta)$, where the sum is
   $L^2$-orthogonal and 
   $\ker(\Delta)=\{ \text{constant functions} \}$, thus 
   $\im(\Delta)=\{f \in C^{\infty}(\Sigma) \;|\; \int_\Sigma f = 0\}$.\zeile
   Moreover, duality and interpolation shows, that $\Delta$ extends to a
   Fredholm operator $H^{s+2}(\Sigma) \longrightarrow H^s(\Sigma)$ of index
   zero for arbitrary $s \in \RR$
\end{proposition}
The class of Fredholm operators is closed under addition of compact
operators and the index remains constant. Thus we easily get the
\begin{corollary}\label{cor.fredholmiso}\ein
   If $\lambda \in C^{\infty}(\Sigma)$ is everywhere non-negative and does
   not vanish identically, then the elliptic operator 
   \begin{equation*}
      L = \Delta + \lambda : 
        H^{s+2}(\Sigma) \longrightarrow H^{s}(\Sigma)
   \end{equation*}
   is an isomorphism for all $s \in \RR$.
\end{corollary}
In particular, this corollary applies to the operator in \eqref{e.lapse} in
cosmological spacetimes.
%
%
%
%
%
%

%%% Local Variables: 
%%% mode: latex
%%% TeX-master: "main"
%%% End: 

%% file: body.tex
%
%
%
%
%
%
\section{Local PMC foliations}\label{s.localpmc}
\subsection{The PMC equations}\label{s.equations}
Let $(M,g)$ be a smooth, globally hyperbolic spacetime with compact Cauchy
surface $\Sigma$. We denote 
the coordinates on $\Sigma$ by $x^i$ and define $U$ to be a Gaussian
neighbourhood of the form 
$I \times \Sigma \supset \;]-\epsilon,\epsilon[\; \times \;\Sigma$, 
with some $\epsilon > 0$ of $\Sigma:=\{t=0\}$. Then we have  
$g(\d_0,\d_0)=-1, \d_0 \perp \d_i$, $i=1,2,3$ and  $\d_0 := \d_t$. $D$
abbreviates the vector $(\d_i)$ and for an intrinsic spacelike
3-vector $v$ set $|v|^2 := g_{ab}v^av^b$.

Let $0 \in J\subset \RR$ denote an interval. A foliation $\{S_{\tau}\}$
of a neighbourhood $U' \subset U$ of $\Sigma$ in $M$ by spacelike
hypersurfaces is given by a map $\varphi : J \longrightarrow
C^{\infty}(\Sigma,\RR),\; \tau \longmapsto \varphi_{\tau}$, with
$|D\varphi_{\tau}| < 1$, $\tau \longmapsto \varphi_{\tau}(x)$
strictly monotone for all $x \in \Sigma$ and
$S_{\tau}=\{(t,x)\,|\,\varphi_{\tau}(x)-t=0\}$. The foliation defines a
unit normal $n$, the first and second fundamental forms $h$ and $k$
respectively and the mean curvature $H$ on each leaf, parametrized by
$\tau$. The coordinates 
on each leaf will be identified with the coordinates on $\Sigma$ by
the action of $\phi_{\tau}^*$, with
$\phi_{\tau}(x)=(\varphi_{\tau}(x),x)$, so we can regard functions on the
leaves as functions on $\Sigma$, parametrized by $\tau$. With these
conventions the foliation fulfills the equation
\begin{subequations}\label{e.pmcfoliation}
\begin{equation}\label{e.foliation}
   \frac{\d \varphi}{\d \tau} = \left( 1 - |D \varphi|^2
   \right)^{\frac{1}{2}} N
   \komma
\end{equation}
where we dropped the subscript $\tau$ from $\varphi$, now simply
considered as a function of $(\tau,x)$. We fix the foliation by
imposing a condition for the lapse function $N$, which has to obey  
\begin{gather}
   L_{\tau} N := \left( \Delta_{\tau} + \lambda_{\tau} \right)\, N = 1 
   \komma \notag \\
   \Delta_{\tau} := -h^{\alpha\beta}\nabla_{\alpha}\nabla_{\beta}
   \komma \label{e.lapsepmc} \\
   \lambda_{\tau} := k_{\alpha\beta}k^{\alpha\beta} + 4 \pi
          (\rho + \tr S)
   \notag \komma
\end{gather}
\end{subequations}
which is the usual lapse equation \eqref{e.lapse} on $S_{\tau}$ (indicated
by the subscript $\tau$ in the equations above), together
with the \emph{prescribed mean curvature condition}
$N n^{\alpha} \d_{\alpha} H = 1$. Thus we force the mean curvature to grow
along the normal vector field of the leaves as described in the
introduction. We define a PMC 
foliation to be a solution of equations \eqref{e.foliation}, 
\eqref{e.lapsepmc} and our goal is to prove uniqueness and local in time
existence of solutions of that system of equations.

To cast the hyperbolic part of the equations into first order symmetric
hyperbolic form, we set 
$w := \tbinom{u}{v} := \tbinom{\varphi}{D\varphi}$ and
express all functions in terms of $w$ and
$N$. Then one finds $n=n(w)$, $h=h(w)$, $k=k(w,Dw)$ and the equations
read
\begin{subequations}
   \label{e.fl}
\begin{gather}
   \d_{\tau}w + A^i(w,N)\,\d_i w + B(w,N,DN) = 0 
   \label{e.f} \\
   L_{\tau}(w,Dw)\,N = 1
   \label{e.L}
\komma
\end{gather}
\end{subequations}
with symmetric $4 \times 4$ matrices $A^i$ and $B$ takes values in $\RR^4$.

Initial values are given by a function $w_0 \in H^s(\Sigma,\RR^4)$,
with $|v_0|<1$, which represents a spacelike hypersurface of
regularity $H^s$ in $M$, with an appropriate choice of $s$. Since we want
to have $S_0=\Sigma$ we set $w_0 = 0$ throughout this work, but we do not
make use of this explicit setting in the sequel. Rather we proceed in full
generality assuming only $w_0 \in H^s(\Sigma,\RR^4)$, spacelike.\zeile
In order
to get a well-posed initial value problem for the system \eqref{e.fl}, we
have to ensure, that $L_0(w_0,Dw_0)$ is an isomorphism of
Sobolev spaces, so that \eqref{e.L} has a unique solution $N_0$ on $w_0$.
In view of corollary \ref{cor.fredholmiso} we require $\lambda_0 \ge 0$,
$\lambda_0 \neq 0$ on $w_0$ which turns 
out to be satisfied for almost all Cauchy surfaces $\Sigma$ in spacetimes
obeying the strong energy condition $\rho+\tr S \ge 0$, while
$\lambda \equiv 0$ would force $\Sigma$ to be maximal and time
symmetric. In the latter case a small deformation of $\Sigma$ would
suffice. Therefore, from now on we focus on cosmological spacetimes in the
sense described in section \ref{s.spacetimes}.
 
The domain $G$ of definition for the coefficients of the equations can be
formally decomposed as 
$G = G_w \times G_{Dw} \times G_N \times G_{DN}$,
where the factors are constraint by the requirements $\abs{v}<1$ in $\RR^3$
and $N>0$ in $\RR$.

To prove local in time existence and
uniqueness for solutions of the system \eqref{e.fl}, equivalent to
\eqref{e.pmcfoliation}, we invoke the following procedure:
We define iteratively a sequence of uncoupled linear equations
(see \eqref{e.flj}), which 
have smooth solutions $(w^j,N^j)$ on non-empty time intervals
$J^j$. Then we 
have to show convergence $(w^j,N^j) \longrightarrow (w,N)$ in some
appropriate Sobolev space on a non-empty time interval $J$. Finally we 
have to verify, that $(w,N)$ represents indeed an unique solution of the
system \eqref{e.fl}.

Let us start with the definition of the sequence mentioned above. 
Denote by $(w^j_0)$ a sequence of smooth functions on
$\Sigma$, with $|v^j_0|<1$ and $w^j_0 \longrightarrow w_0$ in
$H^s(\Sigma)$. The iteration will be defined inductively:
\begin{description}
\item[0)] Set
   \begin{gather*}
       w^0(\tau,x) := w^0_0(x) \quad \forall_{(\tau,x)\in I \times
         \Sigma} \\
       N^0 := 1 \quad \text{on} \quad I \times \Sigma
       \komma
   \end{gather*}
   then one has on $J^0 := I$: $|v^0|=|v^0_0|<1$, which establishes
   $w^0$ as a well-defined family of spacelike hypersurfaces in $U$. Moreover,
   $N^0>0$ by construction and $0 \neq \lambda^0(w^0,Dw^0) \ge 0$, by
   $0 \neq \lambda_0(w_0,Dw_0) \ge 0$ and continuity.   
\item[j)] Now we have a non-empty time interval $J^{j-1}$ for which a
   family of spacelike hypersurfaces $w^{j-1}$ with $0 \neq
   \lambda^{j-1} \ge 0$ is defined, 
   and $N^{j-1}>0$ is given on $J^{j-1}$. Then one gets smooth
   solutions $(w^j,N^j)$ on $J^{j-1}$ for the (linear)
   system
   \begin{subequations}\label{e.flj}
   \begin{equation}\begin{split}
         & \d_{\tau}w^j + A^i(w^{j-1},N^{j-1})\,\d_i w^j +
         B(w^{j-1},N^{j-1},DN^{j-1}) = 0 \komma \\
         & w^j(0) = w^j_0
   \end{split}\label{e.fj}\end{equation}
   \begin{equation}
      L_{\tau}(w^{j-1},Dw^{j-1})\,N^j = 1 \komma \forall_{\tau \in J^{j-1}} 
      \komma
   \label{e.Lj}\end{equation}
   \end{subequations}
   since the above conditions on $w^{j-1}$, $\lambda^{j-1}$ and
   $N^{j-1}$ ensure the regularity of the equations and the
   bijectivity of $L_{\tau}(w^{j-1},Dw^{j-1})$.\zeile
   Set 
   \[
     J^j := \{ \tau \in J^{j-1} \,|\, |v^j|<1,\, 0 \neq \lambda^j \ge 0,\,
     N^j>0\}
     \komma
   \]
   then $J^j \supsetneq \{0\}$ holds: The compactness of $\Sigma$ and
   $|v^j_0|<1$, $0 \neq \lambda^j_0 \ge 0$ give $|v^j|<1$, $0 \neq
   \lambda^j \ge 0$ on a
   non-empty interval by continuity and $N^j>0$ follows from the
   minimum principle for $-L_{\tau}(w^{j-1},Dw^{j-1})N^j = -1$. 
   Hence we have constructed the desired sequence of solutions 
   $(w^j,N^j) \in G$ on $J^j$.
\end{description}
\subsection{The energy estimate}
Now we are going to investigate the convergence properties of
$(w^j,N^j)$. 
Choose $R=(R_1,R_2)$, $R_1,R_2 > 0$ appropriate, such that
\begin{gather*}
   \tilde{G}_w := B_{R_1}(\im w_0) \subset\subset G_w \\
   \tilde{G}_{Dw} := B_{R_2}(\im Dw_0) \subset\subset G_{Dw}
\end{gather*}
holds. Moreover, there are domains $\{1\} \subset \tilde{G}_N
\subset\subset G_N$, $\{0\} \subset \tilde{G}_{DN} \subset\subset
G_{DN}$, which depend on $R$, $||w_0||_{H^s}$. They
will be specified later. 

After this remarks let us turn to the fundamental energy estimate, which
will set us in the position to prove local existence later on (in
subsection \ref{s.localex}).
\begin{lemma}\label{lem.energy}\ein
   Let $s > \tfrac{3}{2}+2$.\zeile
   Then there is a $T>0$, with $J := [-T,T] \subset J^j
   \,,\forall_j$, and a $K>0$, such that for all $j$ and all $\tau \in
   J$:
   \begin{equation}\tag{a}
      \begin{gathered}
         ||w^j(\tau)||_{H^s} \le K \\
         ||N^j(\tau)||_{H^{s+1}} \le K
      \end{gathered}
   \end{equation}
   \begin{equation}\tag{b}
      \im\left(w^j(\tau),Dw^j(\tau),N^j(\tau),DN^j(\tau)\right)
      \,\subset\, 
      \tilde{G}_{w} \times \tilde{G}_{Dw} \times
      \tilde{G}_{N} \times \tilde{G}_{DN} 
      \punkt
   \end{equation}
%   $T$ and $K$ depend only on $R$ and $||w_0||_{H^s}$.
\end{lemma}
\begin{remark*}\ein
   For the reader familiar with energy estimates for quasilinear symmetric
   hyperbolic systems I will explain the connection to the standard
   proofs (see \cite{ra}). 
   For the hyperbolic part of the equations most
   elements of the standard proof carry
   over to the present situation. The difference is due to the coupling to
   the elliptic 
   equation through the dependence on $(N^{j-1},DN^{j-1})$ in \eqref{e.fj}
   and $(w^{j-1},Dw^{j-1})$ in \eqref{e.Lj}. Assuming $s > \tfrac{3}{2}+2$,
   instead of demanding only $s > \tfrac{3}{2}+1$ decouples the estimates
   and the proof splits into the standard proof for the hyperbolic part and
   in a part for the elliptic equation.
\end{remark*}
\begin{proof} [Proof (induction over $j$)]\zeile 
%
% Umschaltung auf \tt-Stil?
\ifraw
   \tt
\fi
%
%
%
%  j = 0
%
   \rule{0em}{1em}\zeile
   For $j=0$ set $T := T^0 := \tfrac{1}{2} \diam J^0$.\zeile
   Since $w^0(\tau)=w^0_0$ we have
   \begin{equation*}
      \cnorm{w^0(\tau)-w_0}{1} = \cnorm{w^0_0-w_0}{1} \le
      C \hnorm{w^0_0-w_0}{s} \le C \epsilon
   \end{equation*}
   by assumption and the Sobolev embedding theorem
   \ref{cor.sobolev}. Without loss of generality we can choose the 
   approximating sequence of the initial values, such that the
   following conditions hold:
   \begin{gather*}
      \epsilon \le \frac{1}{2} \min\left( \frac{R_1}{C}, \frac{R_2}{C}
      \right) \\
      \hnorm{w^j_0}{s}^2 \le 2 \hnorm{w_0}{s}^2 
      \qquad\text{and}\qquad 
      \hnorm{w^{j+1}_0-w^j_0}{s}^2 \le \frac{1}{2^j}
      \qquad\forall_j
   \end{gather*}
   (we will use the two latter conditions for the local existence
   proof in \ref{s.localex}). Statement (b) of the
   lemma follows then immediately with the remark $N^0 \equiv 1$, $DN^0 \equiv
   0$.\zeile
   Next, we have $\hnorm{N^0(\tau)}{s+1}=\hnorm{1}{s+1}=\lnorm{1}{2} \le C$,
   so, if we set
   \begin{equation*}
      K := K^0 := \max \left(\, \hnorm{w_0}{s}+\epsilon, \lnorm{1}{2} \,\right)
   \end{equation*}
   then (a) follows and we have completed the proof for $j=0$.

%
%
%  j-1 ---> j
%
%
%
%
%
%

\vspace{0.3cm}
\hspace*{5cm}\hrulefill\hspace*{5cm}
\vspace{0.75cm}

%
%
%
%
%
%   \rule{0em}{1em}\zeile
   Now we want to perform the induction from $j-1$ to $j$. 
   This proceeds in four steps.\zeile
   Let $\tilde{T}$,$\tilde{K}$ be defined for
   $(w^0,N^0),\dots,(w^{j-1},N^{j-1})$.
   \vspace{1em}\zeile
   \textbf{Step 1 }(\emph{Estimate of $\hnorm{w^j(\tau)}{s}$})\zeile
   If $\alpha$ denotes a multiindex of order 
   $|\alpha|=s$, then differentiation of \eqref{e.fj} yields
   \begin{equation*}\begin{split}
      0 =\dtau \Dalpha \wj 
        & +\underbrace{ A^i(\wjm,\Njm)\,\d_i\Dalpha\wj }_{\text{I}} \\
        & +\underbrace{\left[
            \Dalpha \left( A^i(\wjm,\Njm)\,\d_i\wj \right)
            - A^i(\wjm,\Njm)\,\Dalpha\d_i\wj \right]}_{\text{II}} \\
        & +\underbrace{
               \Dalpha \left( B(\wjm,\Njm,D\Njm) \right)}_{\text{III}}
      \punkt
   \end{split}\end{equation*}
    Now we have to estimate the marked terms. Note, that for $|\tau|\le
    \tilde{T}$ by induction hypothesis all arguments lie in a set whose
    compact closure is contained in $G$. Furthermore, the arguments
    are bounded in the $H^s$, resp. $H^{s+1}$ norms, independent of $\tau$
    and $j$. Then the Moser estimates (prop. \ref{prop.moser}) can be
    applied and one gets for each $|\tau| \le \tilde{T}$
    \begin{equation*}\begin{split}
          \int_{\Sigma}<\Dalpha\wj,\text{I}> 
             & = \int_{\Sigma}<\Dalpha\wj,A^i\,\d_i\Dalpha\wj>
               = \int_{\Sigma}<\Dalpha\wj,(\d_iA^i)\Dalpha\wj> \\
             & \le \lnorm{DA}{\infty} \hnorm{\wj}{s}^2 
               \le C \hnorm{\wj}{s}^2
    \end{split}\end{equation*}
    \begin{equation*}\begin{split}
          \lnorm{\text{II}}{2}
             & \le C \left(\, \lnorm{D^sA}{2}\lnorm{D\wj}{\infty} 
                          + \lnorm{DA}{\infty}\lnorm{D^{s-1}D\wj}{2}
                     \,\right) \\
             & \le C (1+\tilde{K}) 
                   \left(\, \cnorm{\wj}{1} + \hnorm{\wj}{s} \,\right)
    \end{split}\end{equation*}
    \begin{equation*}
       \lnorm{\text{III}}{2} \le C (1+\tilde{K})
       \komma
    \end{equation*}
    hence 
    \begin{equation*}\begin{split}
       \frac{d}{d\tau} \int_{\Sigma} <\Dalpha\wj,\Dalpha\wj>
          & = 2 \int_{\Sigma} <\Dalpha\wj,\dtau\Dalpha\wj> \\
          & \le 2 \int_{\Sigma} <\Dalpha\wj,\text{I}> 
              + 2 \int_{\Sigma} |\Dalpha\wj| | \text{II}+\text{III}| \\
          & \le C (1+\tilde{K}) 
              \left(\, 1 + \cnorm{\wj}{1} + \hnorm{\wj}{s} \,\right)
              \hnorm{\wj}{s}
        \punkt
    \end{split}\end{equation*}
    Integration yields
    \begin{equation*}
       \hnorm{\wj(\tau)}{s}^2 \le \hnorm{\wj_0}{s}^2 + C (1+\tilde{K}) 
         \!\int_0^{|\tau|}\!\!\left(\, 1 + \cnorm{\wj(t)}{1} +
           \hnorm{\wj(t)}{s} \,\right) \hnorm{\wj(t)}{s} \,dt
%       \,.
    \end{equation*}
    This estimate remains true, if we replace $|\tau|$ by
    $\tilde{T}$, $\hnorm{\wj_0}{s}^2$ by $2\hnorm{w_0}{s}$,
    $\cnorm{\wj(t)}{1}$ by $C\hnorm{\wj(t)}{s}$ (Sobolev embedding theorem,
    corollary \ref{cor.sobolev}) and terms of the form $(1+a)^2$ with
    $C(1+a^2)$, so we get 
    \begin{equation*}
       \hnorm{\wj(\tau)}{s}^2 \le 2\hnorm{w_0}{s}^2 + C (1+\tilde{K}) 
         \int_0^{\tilde{T}} \left(\, 1 + \hnorm{\wj(t)}{s}^2 \,\right) \,dt
       \punkt       
    \end{equation*}
    Gronwall (prop. \ref{prop.gronwall}) then gives the final estimate
    \begin{equation*}
       \hnorm{\wj(\tau)}{s}^2 \le \left(\, 1 + 2 \hnorm{w_0}{s}^2 \,\right)
         e^{C(1+\tilde{K})\tilde{T}} \komma
       \qquad \forall_j\;\forall_{|\tau|\le\tilde{T}}
       \punkt
    \end{equation*}
    Set now
    \begin{gather*}
       K_w := \max \left(\, K^0, (1+2\hnorm{w_0}{s}^2)e^C \,\right) \\
       \tilde{T}_w := \min \left(\, T^0, \frac{1}{1+K_w} \,\right)
       \komma
    \end{gather*}
%    then we have $0<K_w=K_w(R,\hnorm{w_0}{s})$,
%    $0<\tilde{T}_w=\tilde{T}_w(R,\hnorm{w_0}{s})$ and for
    then we have $0<K_w$,
    $0<\tilde{T}_w$ and for
    all $j$ and $|\tau|\le\tilde{T}_w$ $\hnorm{\wj(\tau)}{s}$ is
    bounded by $K_w$, hence we have statement (a) of the lemma with
    respect to $w$.
    \vspace{1em}\zeile
    \textbf{Step 2 }(\emph{Boundedness of $\cnorm{\wj(\tau)}{1}$})\zeile
    The equation \eqref{e.fj} gives an Sobolev estimate for $\dtau\wj$:
    \begin{equation*}
       \hnorm{\dtau\wj}{s-1} \le \hnorm{A^i\d_i\wj}{s-1} + \hnorm{B}{s-1} 
                             \le C (1+K_w)^2
       \komma
    \end{equation*}
    since $\hnorm{A^i\d_i\wj}{s-1}\le C(1+K_w)K_w$ and 
    $\hnorm{B}{s-1}\le C(1+K_w)$.
    Further, with multiindices $\beta$,$\gamma$, $|\beta|=s-1$,
    $|\gamma|=s-2$ the differentiated equation yields
    \begin{equation*}
       \lnorm{D^{\gamma}\dtau D\wj}{2} 
         = \lnorm{\dtau D^{\beta}\wj}{2}
         \le \lnorm{\text{I}}{2} + \lnorm{\text{II}}{2} +\lnorm{\text{III}}{2}
         \komma
    \end{equation*}
    with $\alpha$ replaced by $\beta$ in
    each of the terms I,II,III defined above. Since we have
    $s>\tfrac{3}{2}+2$, we can estimate II and III as before with $s$ 
    replaced by $s-1$, and for I we have 
    \begin{equation*}
       \lnorm{\text{I}}{2} 
         = \int_{\Sigma} <A^i\d_iD^{\beta}\wj,A^i\d_iD^{\beta}\wj>
         \le C \lnorm{A}{\infty} \hnorm{\wj}{s}^2
         \le C K_w^2
       \komma
    \end{equation*}
    and together
    \begin{equation*}
       \hnorm{\dtau D\wj}{s-2} 
         \le C K_w^2 + C (1+K_w)(1+K_w)
         \le C (1+K_w)^2
       \punkt
    \end{equation*}
    It follows
    \begin{gather*}
       \begin{split}
          \lnorm{\wj(\tau)-w_0}{\infty}
            & \le \lnorm{\wj_0-w_0}{\infty}
                 + \int_0^{|\tau|} \lnorm{\dtau\wj(t)}{\infty}\,dt \\
            & \le \frac{1}{2} \min(R_1,R_2) + C(1+K_w)^2 |\tau|
       \end{split} \\
       \begin{split}
          \lnorm{D\wj(\tau)-Dw_0}{\infty}
            & \le \lnorm{D\wj_0-Dw_0}{\infty}
                 + \int_0^{|\tau|} \lnorm{\dtau D\wj(t)}{\infty}\,dt \\
            & \le \frac{1}{2} \min(R_1,R_2) + C(1+K_w)^2 |\tau|   
          \punkt
       \end{split}
    \end{gather*}
    If we set
    \begin{equation*}
       T_w := \min \left(\, 
         \tilde{T}_w, \frac{ \frac{1}{2} \min(R_1,R_2) }{C(1+K_w)^2}
                   \,\right)
       \komma
    \end{equation*}
%    then we have $0<T_w=T_w(R,\hnorm{w_0}{s})$ and for
    then we have $0<T_w$ and for
    all $|\tau|\le T_w$ $\cnorm{\wj(\tau)-w_0}{1}$ is bounded by
    $\min(R_1,R_2)$, which means, that statement (b) of the lemma is
    fulfillled with respect to $w$. 
    \vspace{1em}\zeile
    \textbf{Step 3 }(\emph{Estimate of $\hnorm{\Njt}{s+1}$})\zeile
    We are looking for a bound of 
    $\norm{\Ltjm^{-1}}$ in the $H^{s-1}-H^{s+1}$ operator norm. 
    $\Ltjm$ is a sum of terms of the form
    $a_{\tau}^{j-1} D^{\kappa}$, with $\abs{\kappa} \le 2$ and
    $a_{\tau}^{j-1} = a_{\tau}(\wjm,D\wjm)$ where $a_{\tau}$ is a
    smooth function of its arguments and the parameter $\tau$.
 
    We proceed in several steps.\zeile
    At first we consider $\Ltjm$ as an
    operator $H^2(\Sigma) \longrightarrow H^0(\Sigma)=L^2(\Sigma)$
    only. We already know $\norm{L_0(w^0,Dw^0)^{-1}} \le C$ and show
    successively 
    \begin{enumerate}[1)]
    \item $\norm{L_0(\wj,D\wj)^{-1}} \le C \quad\forall_j$
    \item $\norm{\Ltj^{-1}} \le C \quad\forall_j\,
                                   \forall_{\abs{\tau} \le T_N}$,
          with some $T_N > 0$ chosen appropriately.
    \end{enumerate}
    The proof of these statements relies on the following standard fact:
    \begin{note*}\ein
       If $T:H^2(\Sigma) \longrightarrow H^0(\Sigma)$ is invertible
       and $S:H^2(\Sigma) \longrightarrow H^0(\Sigma)$ satisfies
       \begin{equation*}
          \norm{T-S} < \norm{T^{-1}}^{-1}
          \punkt
       \end{equation*}
       Then $S = T - (T-S) = T (1-T^{-1}(T-S))$ is invertible with
       norm 
       $\norm{S^{-1}} \le \norm{T^{-1}} \sum_{k=0}^{\infty}q^k
                      = \tfrac{1}{1-q}\norm{T^{-1}}$, where 
       $q = \norm{T^{-1}(T-S)} < 1$. 
    \end{note*}
    Proof of 1):\ein
      $L_0(w,Dw)^{-1}$ exists by assumption. Now define 
      $\dj{L_0} := L_0(w,Dw) - L_0(\wj,D\wj)$. We want to show 
      $\norm{\dj{L_0}} \le \epsilon$ for an 
      $0 < \epsilon < \norm{L_0(w,Dw)^{-1}}^{-1}$. The note then completes
      the claim.\zeile
      Since $w_0^j \longrightarrow w_0$ in $H^s(\Sigma)$ we can choose 
      the sequence such that for the coefficients of $\dj{L_0}$
      $\hnorm{\dj{a_0}}{s-1} \le \epsilon$ holds (where the 
      Moser estimates (see prop. \ref{prop.moser}) have been applied on some
      expression, coming from 
      the mean value theorem, compare the note in subsection \ref{s.localex}.
      In particular we 
      have for $f \in H^2(\Sigma)$ the estimate
      $\hnorm{\dj{L_0}f}{0} \le C\lnorm{\dj{a_0}}{\infty}\hnorm{f}{2}
                            \le C\hnorm{f}{2}$                       
      and we are done.

    Proof of 2):\ein
      We have shown in 1), that 
      $\norm{L_0(\wj,D\wj)^{-1}} \le \tilde{C}$, hence
      $\tilde{C}^{-1} \le \norm{L_0(\wj,D\wj)^{-1}}^{-1}$.
      We set $\dt{\Lj} := L_0(\wj,D\wj) - \Ltj$ and must show, that
      the estimate $\norm{\dt{\Lj}} < \tilde{C}^{-1}$ holds.\zeile
      For $f \in H^2(\Sigma)$ and $\abs{\kappa} \le 2$ we have
      $\hnorm{\dt{a^j D^{\kappa}f}}{0} \le 
         C\lnorm{\dt{a^j}}{\infty}\hnorm{f}{2}$, 
      and it is enough to find a bound for
      $\lnorm{\dt{a^j}}{\infty}$. 
      For this we mention, that 
      $\dt{a^j}=\int_0^{\tau} \dtau a^j$. Thus
      \begin{equation*}\begin{split}
         \lnorm{\dt{a^j}}{\infty} 
           & \le \lnorm{\max_{\tau} \abs{\dtau a^j}\,\tilde{T}}{\infty}
             = \max_x (\max_{\tau} \abs{\dtau a^j})\, \tilde{T} \\
           & = \max_{\tau} (\max_x \abs{\dtau a^j})\, \tilde{T}
             =   \max_{\tau} \lnorm{\dtau a^j}{\infty}\, \tilde{T}
             \le C(1+K_w) \tilde{T}
      \end{split}\end{equation*}
      Finally choose some $0<T_N\le \tilde{T}$ with  
      $C(1+K_w)T_N \le \tilde{C}^{-1}$.
      
    \vspace{-1ex}\aus
    Until now we have proven
    $\hnorm{\Njt}{2} \le C \norm{\Ltjm^{-1}} \le C$, $\forall_j$, 
    $\forall_{\abs{\tau}\le T_N}$, where $\Ltjm$ acts
    as an operator $H^2(\Sigma) \longrightarrow H^0(\Sigma)$. 
    What remains to do is to get control in higher norms. 
    For this note, that the coefficients $a_{\tau}^{j-1}$ of $\Ltjm$ 
    have regularity $\hnorm{a_{\tau}^{j-1}}{s-1} \le C(1+\tilde{K})$ and
    therefore $\cnorm{a_{\tau}^{j-1}}{1} \le C(1+\tilde{K})$. Now let us 
    consider 
    \begin{equation*}\begin{split}
          \Ltjm (D\Nj) & = D[\Ltjm\Nj] - (D\Ltjm)\Nj \\
          & = - (D\Ltjm)\Nj
       \end{split}\end{equation*}
    and 
    $\lnorm{(Da_{\tau}^{j-1})D^2\Njt}{2}
       \le \lnorm{Da_{\tau}^{j-1}}{\infty}\hnorm{\Njt}{2} 
       \le C(1+\tilde{K})$. 
    Therefore, the right-hand side of the equation above lies in 
    $L^2(\Sigma)$ and is bounded there independently of $j$ 
    and $\tau$, $\Ltjm$ is an isomorphism 
    $H^2(\Sigma) \longrightarrow L^2(\Sigma)$, thus 
    $D\Njt \in H^2(\Sigma)$, bounded by 
    \begin{equation*}
       \hnorm{D\Njt}{2} \le \norm{\Ltjm^{-1}}\lnorm{(D\Ltjm)\Nj}{2}
                        \le C(1+\tilde{K})
       \komma
    \end{equation*}
    thus $\hnorm{\Njt}{3} \le C(1+\tilde{K})$, and we have  
    control over $\Njt$ with one differentiability order
    increased.\zeile
    Now we iterate this procedure to get estimates on higher
    norms. For multiindices $\beta$, $\gamma$, with 
    $\abs{\gamma}=\abs{\beta}-1$ we find 
    \begin{multline*}
       \Ltjm(D^{\beta}\Nj) \\
           = D\{\Ltjm(D^{\gamma}\Nj)\} - (D\Ltjm)D^{\gamma}\Nj \\
           = \sum_{\kappa+\lambda=\gamma} 
               D^{\kappa}(D\Ltjm) D^{\lambda}\Nj
       \komma
    \end{multline*}
    where successively in each step the previous results has been
    inserted. Since we have a bound at most for
    $\hnorm{a_{\tau}^{j-1}}{s-1}$ we get the highest
    order estimate on $\abs{\beta}=s-1$. Thus we have 
    $D^{s-1}\Njt \in H^2(\Sigma)$, with
    $\hnorm{\Njt}{s+1} \le C(1+\tilde{K})$.
    Setting $K_N := C(1+\tilde{K})$ completes the proof of (a) of
    the lemma with respect to $N$.
    \vspace{1em}\zeile
    \textbf{Step 4 }(\emph{Boundedness of $\cnorm{\Njt}{1}$})\zeile
    Multiplication with the Sobolev constant (compare
    cor. \ref{cor.sobolev}) for  
    $H^{s+1} \hookrightarrow C^1$ yields
    \begin{equation*}
       \cnorm{\Njt}{1} \le \underbrace{ C K_N}_{ =:\, R_N}
       \komma
    \end{equation*}
    so we can set
    \begin{equation*}
       \tilde{G}_N := B_{R_N}(\{1\}) \komma\qquad
       \tilde{G}_{DN} := B_{R_N}(\{0\}) \punkt
    \end{equation*}
    These domains are independent of $\tau$ and $j$ and well defined,
    $\Njt$, $D\Njt$ are contained in $\tilde{G}_N$, $\tilde{G}_{DN}$
    for all $j$,$|\tau|\le\tilde{T}$, giving statement (b) of the
    lemma with respect to $N$ and setting
    \begin{gather*}
       T := \min (T_w, T_N) \le \tilde{T} \\
       K := \max (K_w, K_N)
    \end{gather*}
    finishes the proof of the lemma.
\end{proof}

Inspection of the proof shows, that we have shown more.
In particular we have the
\begin{corollary}\label{cor.energy}\ein
   With $s>\tfrac{3}{2}+2$ and $\hnorm{\Njt}{s+1} \le K_N$ it follows
   \begin{equation*}
      \cnorm{\Njt}{3} \le R_N
      \komma
   \end{equation*}
   hence $\im(D^{\alpha}\Nj) \subset \tilde{G}_N$, for $|\alpha|\le
   3$ and all $j$ and $|\tau|\le T$.
\end{corollary}
and 
\begin{corollary}\label{cor.energy2}\ein
   For all $j$ and $\abs{\tau} \le T$ the operator
   $\Ltj : H^2(\Sigma) \longrightarrow L^2(\Sigma)$ is invertible and
   the estimate
   \begin{equation*}
      \norm{\Ltj^{-1}} \le C
   \end{equation*}
   holds.
\end{corollary}
Moreover, the constants $T$ and $K$ depend only on $R$ and
$\hnorm{w_0}{s}$. 
\subsection{Local existence}\label{s.localex}
The aim of the present subsection is to establish the local in time
existence theorem, by proving convergence of the sequence $(w^j,N^j)$
towards a solution of the system \eqref{e.fl}. Since $\Sigma$ is
compact, we will get spatially global results.

With the notation
\begin{equation*}
   \dj{u} := u^{j+1} - u^j 
\end{equation*}
for any given sequence $(u^j)$
we have to estimate the differences $\dj{w}$ and $\dj{N}$ in
Banach spaces of the form
\begin{equation*}
   \left(\, C^l(J,X), \norm{\cdot}_{C^l(J),X} \,\right) \komma\, 
   \norm{u}_{C^l(J),X} := \sum_{i=0}^l \sup_{\tau\in J} \norm{D^iu}_X \komma\,
   J\subset \tilde{J}:=[-\tilde{T},\tilde{T}]
   \komma
\end{equation*}
where $\tilde{T}$ denotes the time constant of the energy estimate and
$X$ an appropriate Sobolev space. Cauchy sequences with respect
to this topology are uniformly convergent in $\tau$.

As a consequence of the mean value theorem we
\begin{note*}\ein
   For $U\subset\RR^k$, $V\subset\RR^l$ let $F:U\times V
   \longrightarrow \RR^m$ a $C^1$-function.\zeile
   Then there is a continuous function $M$ on $(U\times V)^2$, with 
   $M\equiv M_v \oplus M_u \in \RR^{m\times (k+l)}$, 
   $M_v \in \RR^{m\times k}$, $M_u \in \RR^{m \times l}$,
   such that for all $u_1,u_2\in U$, $v_1,v_2\in V$
   \begin{equation*}
      F(u_2,v_2) - F(u_1,v_1) 
        = M \binom{u_2-u_1}{v_2-v_1}
        = M_v (u_2-u_1) + M_u (v_2-v_1)
   \end{equation*}
   holds.
\end{note*}
This kind of replacement will be used frequently in the sequel. 

After these preparations the equations for the differences will be
derived: Fix $\tau\in \tilde{J}$, then it follows from 
$0 = \dtau\wj + \Aijm\,\d_i\wj + \Bjm 
   = \dtau\wjp + \Aij\,\d_i\wjp + \Bj$:
\begin{equation*}\begin{split}
   0 = \dtau\dj{w} 
      & + \Aij\,\d_i\dj{w} 
        + \tilde{A}^i(\wj,\wjm,\Nj,\Njm)\,\d_i\wj
          \left(\begin{smallmatrix} \djm{w} \\ \djm{N}
          \end{smallmatrix}\right) \\
      & + \tilde{B}(\wj,\wjm,\Nj,D\Nj,\Njm,D\Njm)
          \left(\begin{smallmatrix} \djm{w}\\ \djm{N}\\ \djm{DN}
          \end{smallmatrix}\right)
   \komma
\end{split}\end{equation*}
hence the equation fulfilled by the difference $\dj{w}$ is
\begin{subequations}
\begin{equation}\begin{split}
   0 = \dtau\dj{w} & + \Aij\,\d_i\dj{w} \\
                   & + B(\wj,D\wj,\wjm,\Nj,D\Nj,\Njm,D\Njm)
                       \left(\begin{smallmatrix} \djm{w}\\ \djm{N}\\ \djm{DN}
                       \end{smallmatrix}\right)
\end{split}\label{e.djf}\end{equation}
and $B$ is a smooth function of its arguments.

For $N$ one gets in the same manner from 
$1=\Ltjm\Nj=\Ltj\Njp$:
\begin{equation*}\begin{split}
   0 & = \Ltj\Njp - \Ltjm\Nj \\
     & = \Ltj\Njp\!\!-\!\Ltj\Nj +\!\Ltj\Nj -\!\Ltjm\Nj \\
     & = \Ltj\dj{N} + \left(\, \Ltj-\Ltjm \,\right)\,\Nj \\
     & = \Ltj\dj{N} 
       - f_{\tau}(\wj,D\wj,\wjm,D\wjm,\Nj,D\Nj,D^2\Nj)
         \left(\begin{smallmatrix} \djm{w} \\ \djm{Dw}
         \end{smallmatrix}\right)
   \,,
\end{split}\end{equation*}
hence the equation for $\dj{N}$ is
\begin{equation}
   \Ltj\dj{N} = f_{\tau}(\wj,D\wj,\wjm,D\wjm,\Nj,D\Nj,D^2\Nj)
                  \left(\begin{smallmatrix} \djm{w} \\ \djm{Dw}
                  \end{smallmatrix}\right)
\label{e.djL}\end{equation}
\end{subequations}
and $f_{\tau}$ is a smooth function of its arguments and its parameter
$\tau$. 

Inspection of the last equation yields the important
\begin{lemma}\label{lem.Ntow}\ein
   Let $s > \tfrac{3}{2}+2$.\zeile
   Then there exists a constant $C>0$, such that 
   \begin{equation*}
      \hnorm{\dj{\Nt}}{2} \le C \hnorm{\djm{\wt}}{1} 
      \komma\qquad \forall_j \; \forall_{\tau \in \tilde{J}}
   \end{equation*}
   holds.
\end{lemma}
The proof of this lemma is an easy consequence of the fact, that the
right-hand side of \eqref{e.djL} is an element of $L^2(\Sigma)$ and 
$\Ltj: H^2(\Sigma) \longrightarrow L^2(\Sigma)$ is an isomorphism,
with bounded inverse by corollary \ref{cor.energy2}.
\begin{remark}\label{rem.Ntow}\ein
%\hspace*{-4ex}
   With this lemma at hand, we achieve again a decoupling of our
   estimates. Moreover, Cauchy sequences $(\dj{w})$ in
   $C^0(\tilde{J},H^1(\Sigma))$ imply Cauchy sequences $(\dj{N})$ in  
   $C^0(\tilde{J},H^2(\Sigma))$. 
   Since $\hnorm{\dj{\Nt}}{s+1}$ and $\hnorm{\dj{\wt}}{s}$ are uniformly
   bounded, we get the inequality 
   $\hnorm{\dj{\Nt}}{r} \le C \hnorm{\djm{\wt}}{r-1}$ 
   uniformly in $\tau \in \tilde{J}$ for all $1 \le r < s+1$ by
   interpolation. 
\end{remark}

\vspace{0.3cm}
\hspace*{5cm}\hrulefill\hspace*{5cm}
\vspace{0.75cm}

Now let us turn to the estimate of the sequence $(\dj{w})$ in
$C^0(\tilde{J},H^1(\Sigma))$. Applying $D$ on \eqref{e.djf} we infer
from the energy estimate \ref{lem.energy}, that all coefficients of the
differentiated 
equation are bounded in all relevant norms (independent of $j$ and 
$\tau \in \tilde{J}$).
So we get the following estimate (in the same way as we have
done so far in the energy estimate for equation \eqref{e.fj}).
\begin{align*}
   \frac{d}{d\tau}\lnorm{D\dj{w}}{2}^2 
     & = \frac{d}{d\tau} \int_{\Sigma} <D\dj{w},D\dj{w}>
       = 2 \int_{\Sigma} <D\dj{w},\dtau D\dj{w}> \\
     & \le C \left(\, 
             \lnorm{D\dj{w}}{2}+\lnorm{D\djm{w}}{2}+\hnorm{D\djm{N}}{1}
             \,\right)
             \lnorm{D\dj{w}}{2} \komma \\
     \intertext{hence}
     \frac{d}{d\tau}\hnorm{\dj{w}}{1}^2
     & \le C \left(\,
             \hnorm{\dj{w}}{1}^2+\hnorm{\djm{w}}{1}^2+\hnorm{\djm{N}}{2}^2
             \,\right) \\
     & \le C \left(\,
             \hnorm{\dj{w}}{1}^2+\hnorm{\djm{w}}{1}^2+\hnorm{\djmm{w}}{1}^2
             \,\right)
     \komma
\end{align*}
where lemma \ref{lem.Ntow} has been used in the last step. Integration
yields for all $\tau \in \tilde{J}$ (this means $|\tau|\le \tilde{T}$),
using the abbreviation $\dj{w_0} := w^{j+1}_0-w^j_0$:
\begin{equation*}\begin{split}
   \hnorm{\dj{w}(\tau)}{1}^2
     & \le \hnorm{\dj{w_0}}{1}^2 
         + C \!\int_0^{\tilde{T}}\! (\, \hnorm{\dj{w}(t)}{1}^2+
                                    \hnorm{\djm{w}(t)}{1}^2+
                                    \hnorm{\djmm{w}(t)}{1}^2  
                                 \,)\,dt \\
     & \le \left(\, \hnorm{\dj{w_0}}{1}^2
         + C \int_0^{\tilde{T}} (\, \hnorm{\djm{w}(t)}{1}^2+
                                    \hnorm{\djmm{w}(t)}{1}^2
                                \,)\,dt
           \right) e^{C\tilde{T}}
   \komma
\end{split}\end{equation*}
where the Gronwall estimate (prop. \ref{prop.gronwall}) has been applied in
the second step. Since all operations were independent of $\tau$, we can
improve the estimate to
\begin{equation*}
   \norm{\dj{w}}_{C^0(\tilde{J}),H^1}^2
     \le \left(\, \hnorm{\dj{w_0}}{1}^2
         + C \tilde{T} \;(\, \norm{\djm{w}}_{C^0(\tilde{J}),H^1}^2+
                           \norm{\djmm{w}}_{C^0(\tilde{J}),H^1}^2
                                \,)\;
           \right) e^{C\tilde{T}}
   \,.
\end{equation*}
Choose a $T>0$, such that $CTe^{CT}\le q < 1$, and set
$J:=[-T,T]\subset \tilde{J}$. Then we get
\begin{equation*}
   \norm{\dj{w}}_{C^0(J),L^2}^2
     \le \hnorm{\dj{w_0}}{1}^2 \, e^{CT}
         + q \left(\, \norm{\djm{w}}_{C^0(J),H^1}^2+
                      \norm{\djmm{w}}_{C^0(J),H^1}^2
             \,\right)
   \komma
\end{equation*}
summation over $j=2,\dots,m$ and rearranging of the terms gives
\begin{gather*}
   \sum_{j=2}^m \norm{\dj{w}}_{C^0(J),H^1}^2 \\[-3ex] \qquad\qquad
     \le \frac{1}{1-2q} \left(\!
                          (\sum_{j=2}^m \hnorm{\dj{w_0}}{1}^2\,e^{CT})
                        + q \,(\, \norm{\delta_0w}_{C^0(J),H^1}^2+
                                  2 \norm{\delta_1w}_{C^0(J),H^1}^2
                              )
                        \!\right)
\end{gather*}
Clearly, the right-hand side of this inequality is bounded (by choice
of the approximating sequence of initial values, we have
$\hnorm{\dj{w_0}}{1}^2\le\tfrac{1}{2^j}$, see the $j=0$-step
in the proof of the energy estimate),
independent of $m$, so we have found a majorant for the
left-hand side. Thus the sequence $(w^j)$ is a Cauchy sequence, hence
converges in the Banach space $C^0(J,H^1(\Sigma))$. Moreover, 
standard arguments for the interpolation of Sobolev norms yield
convergence in $C^0(J,H^{s'}(\Sigma))$, for all $s'$, with $1<s'<s$,
since $\dj{w}$ is bounded in $C^0(J,H^{s}(\Sigma))$:
\begin{equation*}
   w := \lim_{j\rightarrow\infty} w^j 
   \quad\text{in}\quad C^0(J,H^{s'}(\Sigma))
   \komma
\end{equation*}
and the Sobolev embedding theorem asserts, that $w$ is a classical
function of regularity $C^0(J,C^2(\Sigma)) \subset C^0(J \times
\Sigma)$.

It remains to show, that $w$ is a solution of \eqref{e.f}. For this purpose we
investigate the sequence $(\dtau\dj{w})$. Equation \eqref{e.djf} yields
\begin{equation*}\begin{split}
   \norm{\dtau\dj{w}}_{C^0(J),H^1}
     & = \norm{\Aij\,\d_i\dj{w} + B(\dots)
               \left(\begin{smallmatrix} \djm{w}\\ \djm{N}\\ \djm{DN}
               \end{smallmatrix}\right)}_{C^0(J),H^1} \\
     & \le C \;\left(\, \norm{\dj{w}}_{C^0(J),H^2}+
                        \norm{\djm{w}}_{C^0(J),H^1}+
                        \norm{\djm{N}}_{C^0(J),H^2}
             \,\right) \\
     & \le C \;\left(\, \norm{\dj{w}}_{C^0(J),H^2}+
                        \norm{\djm{w}}_{C^0(J),H^1}+
                        \norm{\djmm{w}}_{C^0(J),H^1}
             \,\right) 
     \komma                        
\end{split}\end{equation*}
with $C$ again independent of $j$ and $\tau$ by the energy estimate.
This shows, that $(\dtau w^j)$ is a Cauchy sequence in
$C^0(J,H^1(\Sigma))$ and since $\dtau\dj{w}$ is bounded in
$C^0(J,H^{s-1}(\Sigma))$ interpolation defines
\begin{equation*}
   \tilde{w} := \lim_{j\rightarrow\infty} \dtau w^j
   \quad\text{in}\quad C^0(J,H^{s'-1}(\Sigma))
   \komma
\end{equation*}
and the Sobolev embedding theorem guaranties, that $\tilde{w}$ is a
classical function, lying in $C^0(J,C^1(\Sigma)) \subset C^0(J \times
\Sigma)$.\zeile
The convergence of the sequences $(w^j)$ and $(\dtau w^j)$ is by
construnction uniform in $\tau$, hence we have $\tilde{w}=\dtau w$ for
$w \in C^0(J,H^{s'}(\Sigma)) \cap C^1(J,H^{s'-1}(\Sigma)) \subset
C^1(J \times \Sigma)$. Moreover, $\dtau w$ obeys
\begin{equation*}
   \dtau w = \tilde{w} = \lim_{j\rightarrow\infty}\dtau\wj
           = \lim_{j\rightarrow\infty} \!\left(\!
                -\Aijm\,\d_i\wj - \Bjm \!\right)
   \,,
%   \komma
\end{equation*}
with smooth and therefore $C^0(J,H^{s'-1}(\Sigma))$-continuous
$A^i,B$.\zeile
We conclude, that $w$ is a solution of equation \eqref{e.f}, if the sequence
$(\Nj)$ converges in $C^0(J,H^{s'}(\Sigma))$. 

\vspace{0.3cm}
\hspace*{5cm}\hrulefill\hspace*{5cm}
\vspace{0.75cm}

Let us now come to the inspection of the sequence $(\dj{N})$. Lemma
\ref{lem.Ntow} and remark \ref{rem.Ntow} show 
$\norm{\dj{N}}_{C^0(J),H^{s'+1}} \le C \norm{\djm{w}}_{C^0(J),H^{s'}}$ and
it easily follows 
\begin{equation*}
   N := \lim_{j\rightarrow\infty} N^j
   \quad\text{in}\quad 
   C^0(J,H^{s'+1}(\Sigma))
   \punkt
\end{equation*}
The smoothness of the coefficients in $\Ltjm\,$, $\tau \in J$ ensures,
that for all $f \in H^{s'+1}(\Sigma)$, $\Ltjm\,f$ converges in the
$H^{s'-1}$ norm to $L_{\tau}(w,Dw)f$, which means convergence
$\Ltjm \stackrel{j\rightarrow\infty}{\longrightarrow} L_{\tau}(w,Dw)$
in the $H^{s'+1}-H^{s'-1}$ operator norm. From this we can infer
$\Ltjm\Nj \longrightarrow L_{\tau}(w,Dw) N$ uniform with respect to
$\tau$ in $H^{s'-1}(\Sigma)$, since 
\begin{gather*}
     \hnorm{L_{\tau}(w,Dw)N - \Ltjm\Nj}{s'-1} \\
     \qquad \le \norm{L_{\tau}(w,Dw)-\Ltjm}\,\hnorm{\Nt}{s'+1} \\
     \qquad\quad + \norm{\Ltjm}\,\hnorm{\Nt-\Njt}{s'+1}
\end{gather*}
and both summands consist of one bounded and one converging to zero
term. The only possible limit for $L_{\tau}(w,Dw)N$ is $1$, since
$\Ltjm\Nj=1$ for all $j$. Hence $N$ solves equation \eqref{e.L}.\zeile
Furthermore, the Sobolev embedding theorem gives regularity
$C^3(\Sigma)$ for each 
$\Nt$, and since $w(\tau) \in C^2(\Sigma)$, $Dw(\tau)
\in C^1(\Sigma)$, $L_{\tau}(w,Dw)$ acts as an
operator $C^3(\Sigma) \longrightarrow C^1(\Sigma)$, establishing $\Nt$
as a classical solution of \eqref{e.L} for each $\tau \in J$.\zeile
Altogether we have proven the following
\begin{theorem}\label{thm.pmcexistence}\ein
   Using the conventions described in subsection \ref{s.equations} given the
   system \eqref{e.fl} with 
   spacelike initial values $w_0 \in H^s(\Sigma)$, $s>\tfrac{3}{2}+2$ and  
   $0 \neq \lambda_0 \ge 0$, there 
   exist a non-empty interval 
   $J=[-T,T]$ and classical solutions $(w,N)$ of \eqref{e.fl}, satisfying
   \begin{gather*}
      w \in C^0(J,H^{s'}(\Sigma)) \cap C^1(J,H^{s'-1}(\Sigma))
        \cap C^1(J \times \Sigma) \\
      N \in C^0(J,H^{s'+1}(\Sigma)) \cap C^0(J,C^3(\Sigma))
      \komma
   \end{gather*}
   for each $s'\in\RR$ with $1<s'<s$.
\end{theorem}
\subsection{Improving the regularity}
Now we want to get rid of the primed $s$, as well
as to investigate the effect of time derivatives 
on $w$ and $N$, in order to make the results more satisfying.

Standard arguments on symmetric hyperbolic equations (see \cite{ra}) show,
that $w(\tau)$ 
possesses indeed regularity $H^s$, since only the energy estimate and
the convergence of the sequence $(\wj)$ are involved in the
arguments. For $\Nt$ we simply replace $s'+1$ by $s$ and we get
\begin{note}\ein\vspace{-3ex}
   \begin{gather*}
      w \in C^0(J,H^s(\Sigma)) \\
      N \in C^0(J,H^s(\Sigma))
      \komma
   \end{gather*}
   so we have a well-defined initial value mapping $w_0 \longmapsto
   (w(\tau),N(\tau))$, mapping $H^s(\Sigma)$ into itself for all
   $\tau \in J$.
\end{note}

We want to know something about the effect of differentiation with
respect to time. For $\wj$ we have seen, that applying $\dtau$ on $\wj$
decreases spatial differentiability by one. For $\Nj$ we see first,
that in $H^2(\Sigma)$ we get (compare corollary \ref{cor.energy2} and
step 3 in the proof of the energy estimate \ref{lem.energy}): 
$\Nt = \lim_j\Njt$ uniformly in $\tau$, and since 
$\Njt \in H^{s+1}(\Sigma)$ this also holds in $H^{s'+1}(\Sigma)$ by
interpolation. The same conclusion for the sequence 
$(\dtau\Njt)$ holds for $s'+1$ replaced by $s'$, since the
coefficients of the operator $\dtau\Ltjm$ have regularity decreased by 
one. Thus $(\dtau\Njt)$ converges uniformly with respect to $\tau$ in
$H^{s'}(\Sigma)$ and this is enough to show, that
$\dtau \Nt = \lim_j \dtau\Njt$ in $H^{s'}(\Sigma)$:
\begin{note}\ein\vspace{-3ex}
   \begin{equation*}
      N \in C^0(J,H^{s'+1}(\Sigma)) \cap C^1(J,H^{s'}(\Sigma))
   \end{equation*}
\end{note}

Let us combine the previous results. Iteration of the
calculations show, that the effect of time differentiating $w$ or $N$
is a decrease of spatial differentiability order by one and we get 
the
\begin{corollary}\label{cor.pmcregularity}\ein\vspace{-2ex}
   \begin{equation*}
      (w,N) \in \;\bigcap_{l \le s}\; C^l(J,H^{s-l}(\Sigma))
      \komma
   \end{equation*}
   in particular
   \begin{equation*}
      w_0 \in C^{\infty}(\Sigma) 
      \qquad\Longrightarrow\qquad
      (w,N) \in C^{\infty}(J \times \Sigma)
      \punkt
   \end{equation*}
\end{corollary}
\subsection{Uniqueness}
Finally we consider the question of uniqueness. 
Let two classical solutions $(w_a,N_a)$, $a=1,2$ of the system \eqref{e.fl}
with initial value $w_0 \in H^s(\Sigma)$ be given, with regularity
\begin{gather*}
   w_a \in C^0(J,H^s(\Sigma)) \cap C^1(J,H^{s-1}(\Sigma)) \\
   N_a \in C^0(J,H^{s'+1}(\Sigma))
\end{gather*}
for $s>\tfrac{3}{2}+2$, $0<s'<s$.\zeile
Define $\mu := w_1-w_2$, 
$\nu := N_1-N_2$, then $\mu$ obeys the difference equation
\begin{equation*}
   \dtau\mu + A^i(w_1,N_1)\,\d_i\mu + \hat{B}(w_a,Dw_a,N_a,DN_a)
             \left(\begin{smallmatrix}\mu \\ \nu \\ D\nu
             \end{smallmatrix}\right) = 0
   \punkt
\end{equation*}
This is an inhomogeneous linear symmetric hyperbolic equation of the form
$\dtau\mu + A^i\d_i\mu+B\mu=f$, whose coefficients depend after
insertion of the solutions $(w_a,N_a)$ only on $(\tau,x)$ and 
$f(\tau,x)=f(\nu(\tau,x),D\nu(\tau,x))$. The energy estimate for
equation of this type yields (see prop. \ref{prop.standardenergy})
\begin{equation*}
   \hnorm{\mu(\tau)}{s'}^2 
     \le C\left( \hnorm{\mu(0)}{s'}^2 + 
                 \sup_{0\le t \le \tau} \hnorm{f(t)}{s'}^2 \right)
   \komma
\end{equation*}

In the same way $\nu$ fulfills an analogous difference equation, and
can be estimated like
\begin{gather*}
   \hnorm{\nu}{s'+1} 
     \le C \norm{L(w_1,Dw_1)-L(w_2,Dw_2)}
     \le C \hnorm{\left(\begin{smallmatrix}\mu\\ D\mu
                 \end{smallmatrix}\right)}{s'-1}
     \le C \hnorm{\mu}{s'} \\
   \hnorm{D\nu}{s'} 
     \le C \hnorm{\left(\begin{smallmatrix}\mu\\ D\mu\\ D^2\mu
                  \end{smallmatrix}\right)}{s'-2}
     \le C \hnorm{\mu}{s'}
   \komma
\end{gather*}
for each $\tau$. This simplifies the energy estimate for $\mu$ to
\begin{equation*}
   \hnorm{\mu(\tau)}{s'}^2 
      \le C \left( \hnorm{\mu(0)}{s'}^2 + 
                   \sup_{0\le t \le \tau} \hnorm{\mu(t)}{s'}^2 \right)
    \punkt
\end{equation*}
Since we have $\mu(0)=0$ by assumption, we find $\mu(\tau)=0$ for all
$\tau \in J$. 
Furthermore $\hnorm{\nu(\tau)}{s'+1} \le C \hnorm{\mu(\tau)}{s'}$ yields
$\nu=0$, hence
\begin{gather*}
   w_1 = w_2 \\
   N_1 = N_2
   \komma
\end{gather*}
as desired.
%
%
%
%
%
%
%
%
%
%
%

%%% Local Variables: 
%%% mode: latex
%%% TeX-master: "main"
%%% End: 

%% file: conclusion.tex
%
%
%
%
%
\section{Conclusion and outlook}
Putting together theorem \ref{thm.pmcexistence}, corollary
\ref{cor.pmcregularity} and the uniqueness result we get the final
\begin{theorem}\label{thm.pmc}\ein
   Let $(M,g)$ be a smooth, globally hyperbolic spacetime, obeying the
   strong energy condition, with compact Cauchy surface $\Sigma$ and
   \begin{equation*}
      \lambda = \abs{k}^2 + 4 \pi (\rho + \tr S)
   \end{equation*}
   does not vanish identically on $\Sigma$.\zeile
   Then there exists a $T>0$ and a unique smooth PMC foliation
   $\{S_{\tau}\}$,  $\tau \in [-T,T]$ in $(M,g)$, with $\Sigma=S_0$.
\end{theorem}

Note, that the setting here is quite general, no
symmetry assumptions have to be made and essentially the strong energy
condition turned out to be sufficient for the local in time existence of a
unique PMC foliation up to the choice of an initial Cauchy surface. 

What remains to do is to globalize the result. Here are two problems
involved: How large is the interval of values taken by the time coordinate
and does the global foliation then cover the whole spacetime? To answer
these questions in general there seem to be no techniques available up to
now. One 
strategy to obtain global results is, to study first spacetimes with some
spatial symmetry, taking advantage of the simplifications of the
equations. Then the hope is, that the techniques developed in these cases
give insight into the nature of more general classes of spacetimes, by
successively lowering the degree of symmetry. This line of attack will be
the content of a forthcoming paper. 
\vspace{1em}\zeile
\textbf{Acknowledgements:}
I want to thank Alan Rendall, who guided me through this part of my
work done so far in the Max Planck Institute for Gravitational Physics.
%
%
%
%
%

%%% Local Variables: 
%%% mode: latex
%%% TeX-master: "main"
%%% End: 

%% file: bib.tex
%
%
%
%
%
\pagebreak

%
%
%
%
%
%%% Local Variables: 
%%% mode: latex
%%% TeX-master: "main"
%%% End: 